\documentclass[]{aastex631}

\DeclareUnicodeCharacter{2212}{-}

\shorttitle{On the Influence of the solar wind on the propagation of Earth-impacting Coronal Mass Ejections. }
\shortauthors{Kumar et al.}
\graphicspath{{./}{figures/}}
\begin{document}
\title{On The influence Of The Solar Wind On The Propagation Of Earth-impacting Coronal Mass Ejections} 

\author[0000-0002-3902-5526]{Sandeep Kumar}
\affiliation{Udaipur Solar Observatory, Physical Research Laboratory, Udaipur, 313001, India}
\affiliation{Discipline of Physics, Indian Institute of Technology Gandhinagar, Palaj, Gandhinagar-382 355, Gujarat, India}

\author[0000-0002-0452-5838]{Nandita Srivastava}
\affiliation{Udaipur Solar Observatory, Physical Research Laboratory, Udaipur, 313001, India}

\author[0000-0001-5894-9954]{Nat Gopalswamy}
\affiliation{NASA Goddard Space Flight Center, Greenbelt, MD 20771, USA}
\author{Ashutosh Dash}
\affiliation{Central University of Haryana, Jant-Pali, Mahendergarh
Haryana, India}

\begin{abstract}

Coronal Mass Ejections (CMEs) are subject to changes in their direction of propagation, tilt, and other properties as they interact with the variable solar wind. We investigated the heliospheric propagation of 15 Earth-impacting CMEs observed during April 2010 to August 2018 in the field of view (FOV) of the Heliospheric Imager (HI) onboard the STEREO. About half of the 15 events followed self-similar expansion up to 40 $R_\odot$. The remaining events showed deflection either in latitude, longitude, or a tilt change. Only two events showed significant rotation in the HI1 FOV. We also use toroidal and cylindrical flux rope fitting on the in situ observations of interplanetary magnetic field (IMF) and solar wind parameters to estimate the tilt at L1 for these two events. Although the sample size is small, this study suggests that CME rotation is not very common in the heliosphere. We attributed the observed deflections and rotations of CMEs to a combination of factors,
including their interaction with the ambient solar wind and the influence of the ambient magnetic field. These findings contribute to our understanding of the complex dynamics involved in CME propagation and highlight the need for comprehensive modeling and observational studies to improve space weather prediction. In particular, HI observations help us to connect observations near the Sun and near Earth, improving our understanding of how CMEs move through the heliosphere.

\end{abstract}

\keywords{Coronal Mass Ejections, GCS Fitting, Heliospheric Imaging, CME rotation, CME deflection }


\section{Introduction} 
\label{sec:intro}
Coronal mass ejections (CMEs) are magnetized plasmas with high kinetic and magnetic energies in the heliosphere. Understanding the trajectory of CMEs through the heliosphere and their interaction with the surrounding solar wind helps to determine their impact on the Earth. This essentially requires multi-point remote-sensing observations of CMEs.\\
Previous studies of CMEs based on the observations made by the Large Angle and Spectrometric Coronagraph (LASCO) \citep{lasco:1995} field of view (FOV) on board the SOHO mission \citep{soho:1995} suggested that CMEs follow a three-step evolutionary phase \citep{Zhang:2001}. This includes an initiation phase, an impulsive acceleration phase, and a propagation phase. In the propagation phase, most CMEs are expected to maintain a constant direction of propagation in the heliosphere \citep{Vourlidas:2010}. However, in some cases, CMEs are also known to get deflected in the lower corona towards the heliospheric current sheet (HCS) region due to the interaction with the ambient magnetic field of the Sun. Near the solar minimum, the HCS generally lies along the equator; during that time, CMEs are deflected toward the equator due to the bipolar structure of the magnetic field \citep{Macqueen:1986, GP_TP:2000,Filipo:2001,gop:2008JASTP,Paul:2013,kc:2015}. Using a coronal hole influence parameter (CHIP) \citep{gop_mac_2009}, \cite{makela:2013} suggested that CMEs can be deflected by nearby coronal holes towards the polarity inversion line (PIL); they also studied the influence of coronal hole magnetic fields on magnetic cloud (MC) and non-magnetic cloud-associated events. Based on the events with  CHIP values larger than 2.6 G, \cite{makela:2013} reported that non-MC events were deflected away from the Sun-Earth line, whereas MC events were deflected towards the Sun-Earth line. Their finding supported the idea that all ICMEs are MC/flux rope and those events which are observed as non MC at 1 AU were deflected away from Sun-Earth line by nearby coronal holes.
Based on a study of 36 CMEs observed during 2007 and 2010, \cite{jones:2017} found that 28 of the CMEs were deflected toward the HCS region. Their findings also suggest that the magnitude of deflection is directly proportional to the latitudinal distance of the CME from the HCS.

CMEs are also known to deflect in the longitudinal direction due to their interaction with the ambient solar wind. Earlier studies show that eastward heliospheric deflection of fast CMEs propagating in the slow solar wind is due to interaction with Parker spiral in the heliosphere \citep{gos:1987,Wang:2004}. Such deflection leads to an asymmetry in the source region of Earth-impacting fast CMEs around the disk center of the Sun, i.e., the distribution is more biased toward the west limb of the Sun for fast CMEs.
 Apart from the deflection in the latitudinal and longitudinal directions, CMEs are also reported to undergo rotation in the corona and heliosphere. A rapidly rotating CME ($60^\circ$/day) in the lower corona (below 5 $R_\odot$) was reported by \cite{Vlourlidas:2011}. Previous studies have highlighted a significant disparity in the tilt of CMEs estimated near the Sun, inferred from source regions, and the tilt estimated at L1 based on in-situ data \citep{Palemrio:2018,xi:2021}. This observed mismatch suggests a rotation of the CME in the heliosphere. Although a few studies on CME propagation have been reported based on continuous stereoscopic reconstruction of CMEs in the heliosphere \citep{rot:2018,rot:2020,rot:2021,rot:2022}, only a few of them report on the CME orientation, i.e., the tilt in HI1 images \citep{Kumar_2023}.

The continuous tracking and stereoscopic reconstruction can help us bridge the gap between the near-sun observations and the observations at L1. In our previous study, we reported on the continuous rotation of the October 5, 2012, CME by tracking it up to heliospheric distances of 58 $R_\odot$ using HI1 images \citep{Kumar_2023}. A pertinent question following this finding is whether CME rotation is a rare phenomenon or not during its propagation. A single case study of the CME in the heliosphere using HI1 images is not sufficient to understand the complex dynamics and factors affecting CME trajectory. Therefore, it is necessary to extend such a study to several cases to enhance our understanding of CME propagation in the heliosphere.\\
 
In the present work, we analyse 15 geo-effective (Dst $\leq$ -75 nT) CMEs observed during April 2010 to August 2018 using GCS reconstruction in the corona and heliosphere. We continuously track
each CME to study its trajectory in the heliosphere. We investigate and correlate the observed changes in the trajectories of the CME with the ambient
medium of the CME, i.e., solar wind and magnetic field conditions of the ambient medium.
 
\section{OBSERVATIONS AND ANALYSIS}
\label{sec:obs_ana}
\subsection{Data Selection}
 We selected Earth-impacting CMEs that occurred between April 2010 and August 2018 and were associated with geomagnetic storms with a Dst index less than -75 nT. These events were further filtered based on the availability of data from various coronagraphs, i.e., LASCO-C2 $\&$ C3 onboard SOHO and SECCHI/COR2 \citep{secchi:2008} and HI1 \citep{hi_2009} onboard the STEREO spacecraft \citep{kaiser:2008}, allowing maximum possible viewpoints for stereoscopic reconstruction. Finally, we had 15 events, fulfilling the selection criteria. Table~\ref {tab:events_all} provides a summary of these events, including the date and time of the first detection of CMEs in LASCO-C2 images. The source location mentioned in the table is taken from the CDAW Data Center, SOHO/LASCO CME CATALOG (\url{https://cdaw.gsfc.nasa.gov/CME_list/}) \citep{seiji_2004,gp:2009}.
 We used the coronagraphic images from LASCO-C2 $\&$ C3, SECCHI/COR2, and HI1 to extend CME trajectory tracking, in the FOV of HI1, i.e., up to 80 $R_\odot$. We also used the interplanetary magnetic field (IMF) data taken by the Advanced Composition Explorer (ACE) \citep{ace:1998} to model the CME flux rope at L1. We used the standard synoptic magnetic maps of the Sun from the Global Oscillation Network Group (GONG) website (\url{https://gong.nso.edu/data/magmap/crmap.html}) to investigate the ambient magnetic field of the Sun \citep{hill:2018}. We also employed the WSA  model from NASA CCMC \citep{wsa_2003} to estimate the background solar wind velocity at 21 $R_\odot$. \\

\subsection{CME flux rope model in the corona and heliosphere}

The Graduated Cylindrical Shell (GCS) reconstruction \citep{Thernisien_2006} is a framework to illustrate the self-similar expansion of a CME. This model is characterized by two conical legs and a graduated cylindrical shell representing the CME flux rope. This model has six parameters, i.e.,  latitude ($\theta$) and longitude ($\phi$) of the axis, the half angle between conical legs ($\alpha$), the tilt of the model axis with the ecliptic plane ($\gamma$), the axis height ($h$), and the aspect ratio ($\kappa $).\\

The initial two parameters, $\theta$ and $\phi$, estimate the direction of CME propagation. The last four parameters, $\alpha$, $\gamma$, $h$, and $\kappa$, collectively provide a description of the overall geometric configuration of the fitted CME.

We implemented the GCS reconstruction model on the images from the SOHO/LASCO-C2 \& C3, and STEREO/COR2 running difference images to track the CME in the heliosphere using a module in Python\footnote{\url{https://github.com/johan12345/gcs_python}} \citep{johan:2021}. We enhanced the existing Python module to incorporate HI1 level2\footnote{\url{https://stereo-ssc.nascom.nasa.gov/data/ins_data/secchi/secchi_hi/L2_11_25/}} images and were able to continuously track the CME structure in the HI1 FOV. A similar approach was followed by \cite{gop:2022} who incorporated GCS reconstruction in the HI1 image of the August 21, 2018 CME. The continuous tracking of GCS parameters allowed us to effectively fit the model to observational data up to approximately 70 $R_\odot$ (For Event 12 in Table:\ref{tab:events_all}). 

It is important to note that for events until October 2014, we had observations available from three viewpoints up to the LASCO-C3 FOV, i.e., STEREO-A, STEREO-B and SOHO. For these events, we utilized data from both COR2 \& HI1 and LASCO-C2 \& C3. However, for events that occurred after October 2014, we were limited to only two viewpoints up to the LASCO-C3 FOV: SECCHI/COR2 \& HI1 on STEREO-A and SOHO/LASCO-C2 \& C3. For all events beyond LASCO-C3 FOV, we were restricted to observations from the HI1 (STEREO-A and STEREO-B) instruments alone. Moreover, since most of the events are Earth-directed within $\pm 30^\circ$ of the Sun-Earth line, the actual trackable distances using 3D GCS reconstruction of these CMEs in LASCO-C2 \& C3  FOV is larger than the normal FOV of LASCO-C2 \& C3 due to projection effects.

Along with the GCS reconstruction in HI1 FOV, we employed the cylindrical and toroidal flux rope models developed by \cite{Maru:2007, Maru:2017} to fit IMF data at L1 observed by ACE. Both models are characterized as force-free constant-$\alpha$ models, offering the latitude and longitude of the flux rope axis in GSE coordinates. These coordinates, in turn, enable the estimation of the tilt of the axis of the flux rope.
\subsection{Ambient Solar Wind and Magnetic Field Environment}

To understand the ambient magnetic field and solar wind of the CME, we used pfsspy \citep{david_stansby_2019_2566462} for potential field source surface (PFSS) extrapolation of the photospheric magnetic field of the Sun up to  2.5 $R_\odot$; PFSS extrapolation also gives us an idea of the overall structure of the magnetic field even above  2.5 $R_\odot$. We also used Wang–Sheeley–Arge \citep{Arge:2000,wsa_2003} code for solar wind velocity background at 21 $R_\odot$ from NASA/CCMC \footnote{\url{https://ccmc.gsfc.nasa.gov/results/index.php}}. 
\begin{table}[]
\caption{List of 15 Earth-impacting events selected for our study. Event 7 has been studied in detail by \cite{Kumar_2023}. Events with superscripts YL, KG, SP, and NG were reported in \cite{YL:2011}, \cite{kc:2018}, \cite{pat:2016}, and \cite{rot:2022}, respectively. }
    \centering
    \begin{tabular}{|c|c|c|c|c|}
    \hline
     Event  & Date & Time (UT)& Source Location & Dst(nT) \\
    \hline
        $1^{YL}$  &  2010-04-03 & 10:33 & S25E00 &-81 \\
    \hline
        2  &2010-05-23 & 18:06 &  N16W10 &-80 \\

    \hline
        $3^{KG}$ & 2011-09-06 & 23:05 &  N14W18 &-75 \\

       \hline
      $4^{SP}$ &2012-03-07 & 01:30 &  N15E26 &-145 \\

    \hline
 5 &   2012-07-03& 21:25 & N14W23& -78\\
         \hline
$6^{KG}$ & 2012-09-28 & 00:12&  N09W31&  -122\\
  \hline
$7^{*SK}$ & 2012-10-05 & 02:48 & S23W31 & -105\\
  \hline
8 & 2013-06-02 & 16:25 & N14W25 &-78\\
  \hline
9 &  2015-12-28 & 12:12 &S23W11 &  -116\\
  \hline
10 & 2016-01-14 &  23:24 &  S22W11 & -93\\
  \hline
11 & 2016-10-09 & 02:24 & S02E39 & -110\\
  \hline
12 & 2017-05-23 & 05:00 &S03W01 & -125\\
  \hline
13 & 2017-09-04 & 20:36 & S10W16 & -122\\
  \hline
14 & 2017-09-06 & 12:24 & S08W33 & -109\\
  \hline
$15^{NG}$ & 2018-08-20 & 21:24 & N21W08 &-175\\
     \hline
    \end{tabular}
    
    \label{tab:events_all}
\end{table}

 \section{Results and Discussion}

\label{ch:2}
\begin{table}[]
\caption{Summary of the events that showed significant changes in GCS parameters. Here EW refers to Eastward deflection, and EqW is Equatorward deflection based on first independent fitting. The events in red denotes the cases of CME rotation, and the asterisk denotes the event reported in \cite{Kumar_2023}. }
    \centering
    \begin{tabular}{|c|c|c|c|}
    \hline
     Event & First fitted height/ Last fitted height & Change Observed & Velocity (km/s) \\
    \hline
        1  &  6 $R_\odot$ / 49 $R_\odot$  &Long. $10^\circ$ EW  &800 \\
    \hline
        2  &  8 $R_\odot$/ 34$R_\odot$  &Lat. $10^\circ$ EqW   &450 \\

    \hline
        3 & 9 $R_\odot$/54 $R_\odot$    &Long. $11^\circ$ EW    &850 \\

       \hline
        
    6  &    6 $R_\odot$/ 54 $R_\odot$  &Lat. $10^\circ$ EqW  &    850\\
  \hline
\textcolor{red}{7*}  &  7 $R_\odot$/ 58 $R_\odot$ & tilt $21^\circ$ Anti clockwise   &600\\
  \hline

\textcolor{red}{12}  & 10 $R_\odot$/ 70 $R_\odot$ & tilt. $19^\circ$Anti clockwise & 400\\
  \hline
13  &    13 $R_\odot$/ 42 $R_\odot$      &Long. $25^\circ$ EW    & 1800\\
  \hline
14  &     14 $R_\odot$/ 35 $R_\odot$  &Long. $13^\circ$ EW  & 1450\\
  \hline

    \end{tabular}
    
    \label{tab:change_all}
\end{table}

\subsection{GCS Reconstruction}

Each CME selected for our analysis was tracked from LASCO-C2, COR2 and LASCO-C3 FOV  up to the last frame/time in the HI1 images. Considering uncertainties in the GCS fitting, we focused on events that exhibited a change greater than $10^\circ$ in latitudes and longitudes, i.e., difference of initial and final GCS reconstruction. Our threshold for tilt change is almost twice as compared to the latitude and longitude because errors in tilt can be much larger than those in latitude or longitude \citep{Kumar_2023}. The list of selected events that showed significant change is given in Table~\ref{tab:change_all}. All the changes observed and listed for each case in the table were in the same direction, i.e., continuously increasing or decreasing. Note that for each event, we selected the time for the first GCS reconstruction when the CME was fully visible and the flux rope structure was well developed in the COR2 and C2 FOV.  This choice was made to capture the flux rope at its mature stage. The initial height for the GCS reconstruction ranged between 6 and 14 $R_\odot$ for all events. Additionally, the final height of the GCS reconstruction was determined by the time at which the CME structure remained clearly visible and intact in the HI1 FOV. Although some faster CMEs were observable at higher distances, their features became too distorted at those heights to accurately implement the GCS model. We also estimated the average true velocity. Based on the height-time analysis of the 15 Events we conclude the following:

\begin{itemize}
    \item 7 out of 15 CMEs did not exhibit any significant deviations from self-similar expansion, i.e.,  changes in the values of latitude, longitude, or tilt, indicating that CMEs reach a steady self-similar expansion state above $\approx$ 10$R_\odot$ as shown by \cite{Vourlidas:2010}.
    \item Six events displayed significant deflection ($\geq10^\circ$) in either longitude or latitude.
    \item Only two events (October 5, 2012, and May 23, 2017) showed an increase in tilt as shown in red color in Table~\ref{tab:change_all}.
    \item  Two CMEs (Event 2 and 6) were latitudinally deflected toward the equator (EqW). Longitudinal deflections were observed in four events (Events 1, 3, 13 and 14) in the EastWard (EW) direction.
\end{itemize}
As reported in our previous study \citep{Kumar_2023}, GCS parameters have uncertainties associated with them \citep{verbek:2023}. We also found that latitude and longitude vary the least based on the results of four independent fittings \citep{Kumar_2023}. Therefore, to confirm the changes in latitudes and longitudes for 6 Events (2, 6, 1, 3, 13, and 14),  two independent fittings were performed by the coauthors, for the events showing latitudinal and longitudinal deflection: first  fitting was done with a cadence of one hour, (the estimated heights from this fit are shown in Table 2, column 2), the second independent fitting was done for  the first, and last images of the first independent fitting, to confirm the change in the first set of fitting. As a part of the second independent fitting for the 4 cases of longitudinal deflection (Events 1, 3, 13 and 14) and 2 cases of latitudinal deflection (Events 2 and 6), we implemented GCS reconstruction on the CME structure at a time prior to the first fitting based on the well-developed CME feature. From the two independent fittings on well developed CME features,  we found similar results for latitudinal and longitudinal deflecting CMEs from two independent fittings. However, for the case of Event 12, which showed a change in tilt, three independent fittings were performed on images taken at a cadence of 1 hour and all three fittings showed similar results. The estimated changes in the values of tilt  are (i) 21$^\circ$ (24$^\circ$-45$^\circ$), (ii) 19$^\circ$ (13$^\circ$-32$^\circ$) and  (iii) 20$^\circ$ (15$^\circ$-35$^\circ$).
\subsection{Ambient Environment Investigation}
We examined the ambient magnetic field and solar wind speed to understand the possible reasons for the deflection of the CMEs listed in Table~\ref {tab:change_all}. 

\begin{figure}[!h]
    \begin{center}
    \includegraphics[width=19.0cm]{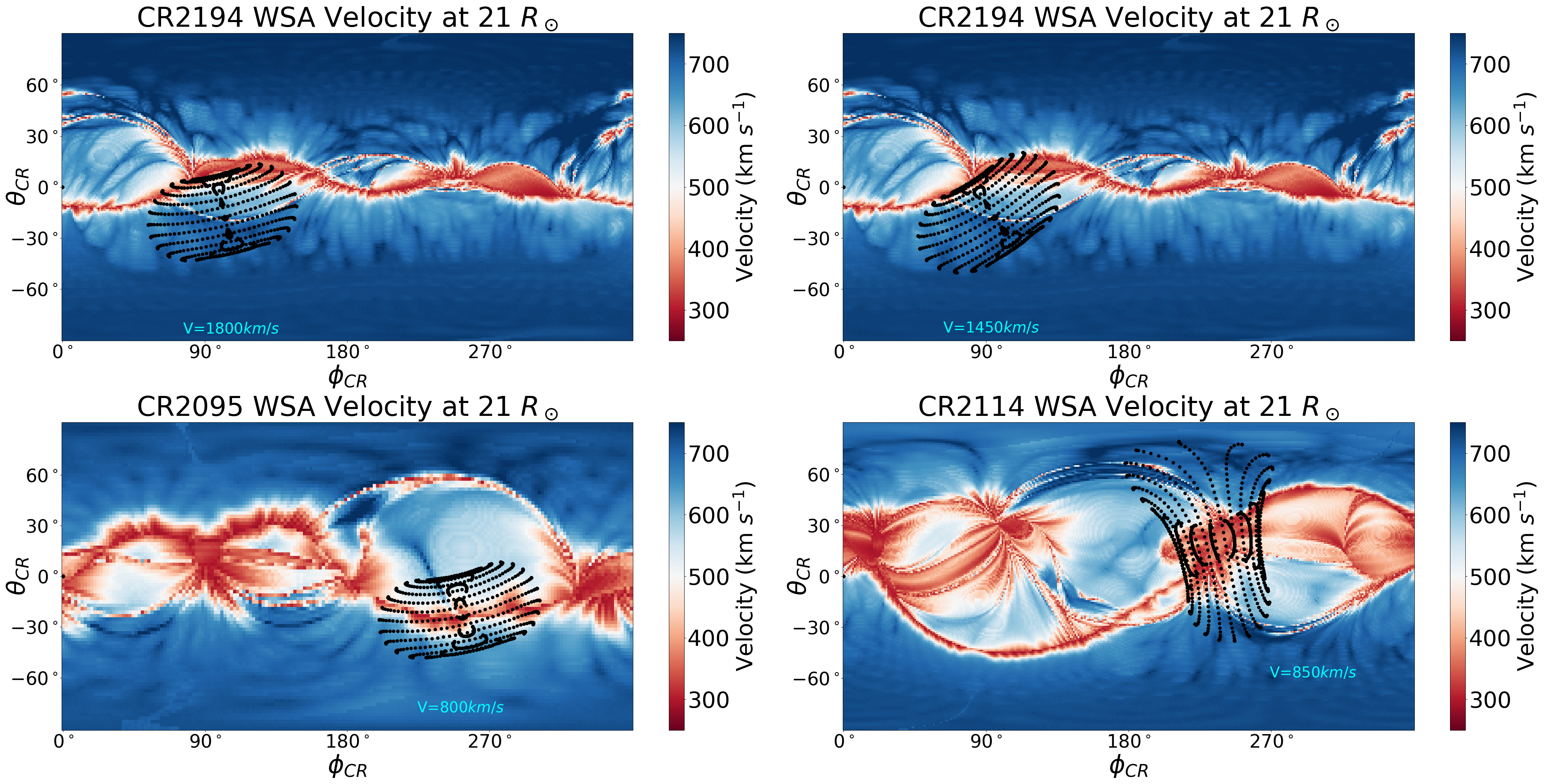} %
    \caption{Solar wind velocity background at 21 $R_\odot$ of the events showing the longitudinal deflection. The top left and right panels show the background for Events 13 and  14, respectively. The bottom left and right panels show the background for Events 1 and 3, respectively.  The black dotted mesh shows the position of the CME and GCS mesh, and velocities are indicated near the mesh. Events 13 and 14 are plotted with a lower half angle and kappa than what was used in GCS fitting to show their propagation direction properly.  $\theta_{CR}$ and $\phi_{CR}$ represent Carrington latitude and longitude respectively.} 
    \label{fig:sol_env}
    \end{center}
\end{figure}

Figure~\ref{fig:sol_env} shows the solar wind background at  $\approx21 R_\odot$ for the longitudinally  deflected events.

\begin{figure}[!h]
    \begin{center}
    \includegraphics[width=18.5cm]{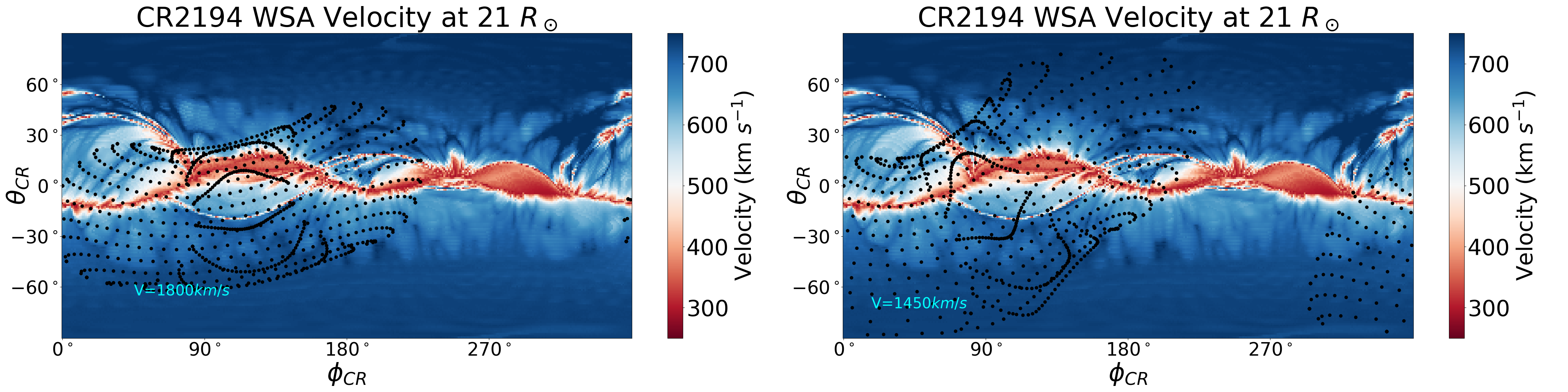}

    \caption{Solar wind velocity background of Events 13 and  14 at 21 $R_\odot$, respectively from left to right,  plotted with the actual fitted parameters of GCS reconstruction.}
    \label{fig:sol_env_rel}
    \end{center}
\end{figure}

All four CMEs (Event 1, 3, 13 and 14) propagated in a slow solar wind relative to their own speeds, as illustrated in Figure~\ref{fig:sol_env}.
In Figure~\ref{fig:sol_env}, Events 13 and 14 are plotted with a smaller half angle and kappa than the actual GCS fitting to show their direction of propagation properly. In Figure~\ref{fig:sol_env_rel}, Events 13 and 14 are re-plotted with the actual GCS parameters to present the real extension of CMEs in the solar wind environment.
The two fast events, 13 and 14, with average speeds of 1800 km/s and 1450 km/s, respectively, occurred within two days and were longitudinally deflected in the eastward direction towards the Sun-Earth line.

The deflection due to solar wind drag depends on the cross-section area of CME and the difference in the speed of CME and ambient solar wind, CME density and mass \citep{cargil:2004}. For CMEs traveling in a similar ambient medium maintaining similar extension/shape in the heliosphere, their relative deflection depends upon their average speeds. Our analysis confirms this. For instance, fast Events 13 and 14, with velocities of 1800 km/s and 1450 km/s  were deflected by $25^\circ$ and $13^\circ$, respectively. They were extremely fast events fitted with higher kappa (0.67 and 0.79 respectively) and half angle ($72^\circ$ and $78^\circ$) in GCS  (see Figure:~\ref{fig:sol_env_rel}). These events propagated in nearly identical solar wind backgrounds as shown in Figure:~\ref{fig:sol_env} (top panel) and in Figure:~\ref{fig:sol_env_rel}. Events 1 and 3, with near identical velocities of 800 km/s and 850 km/s, exhibited similar magnitude of deflection $\approx10^\circ$. Moreover, the dependence between the magnitude of deflection and velocities is evident when we compare Events 1 and 3 (with average velocity $\approx$ 825 km/s and average deflection $10.5^\circ$), with Events 13 and 14  (with average velocity $\approx$ 1625 km/s and average deflection  $19^\circ$).\\
For the faster events, i.e., Events 13 and 14, the solar wind drag is expected to dominate over the magnetic field interaction at lower heights, i.e., 3.4 $R_\odot$ to 4 $R_\odot$ \citep{sach:2017}. These heights are lower than the height at which we begin the first GCS fitting, i.e., $\approx $ 6 $R_\odot$. Therefore, we expect the solar wind drag for Events  13 and 14,  to play a significant role compared to the magnetic field in the domain of GCS reconstruction beyond 6 $R_\odot$.  We investigated the magnetic field environment for Event 1 and 3 (not included in the manuscript). As these events are slower than Events 13 and 14, the ambient magnetic field environment may have influenced their propagation. However, we found that the ambient magnetic field did not influence the CME propagation for Events 1 and 3.

\begin{figure}[!h]
    \begin{center}
    \includegraphics[width=17cm]{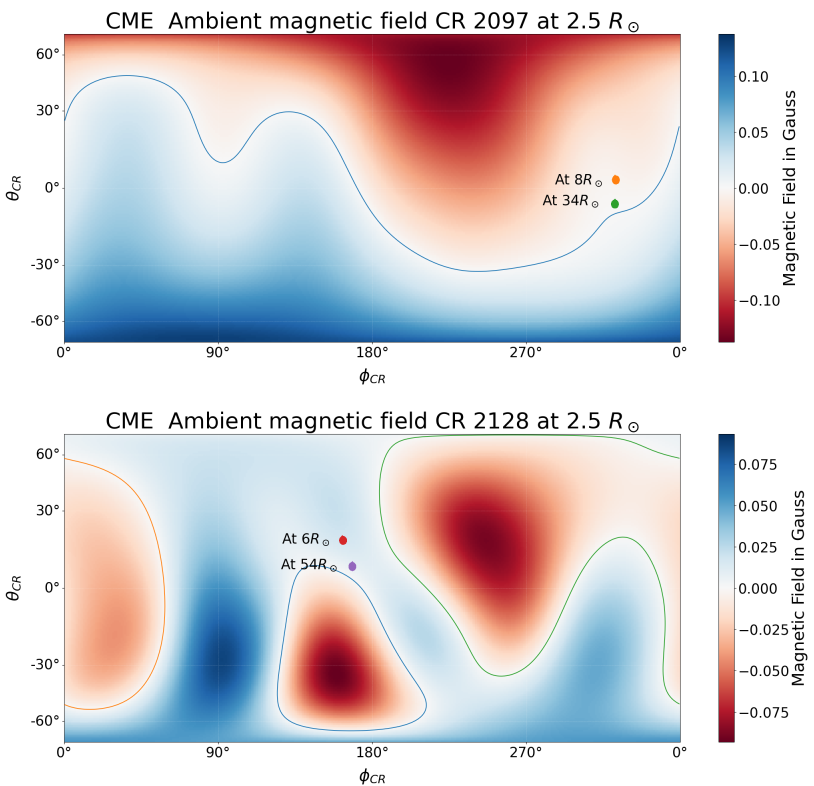} %
   \caption{Magnetic field environment of the events showing the latitudinal deflection inferred from PFSS extrapolation of respective Carrington magnetic field maps. The top panel shows the May 23, 2010 (Event 2), and the bottom panel shows the 28 September 2012 (Event 6) CME. The color bar shows the magnetic field in Gauss. $\theta_{CR}$ and $\phi_{CR}$ represent Carrington latitude and longitude respectively.}
    \label{fig:mag_env}
    \end{center}
\end{figure}

Figure~\ref{fig:mag_env} shows the magnetic field environment (extrapolated using PFSS) of the latitudinally deflected Event 2 as it propagated from 8 $R_\odot$ to 34 $R_\odot$ and Event 6 as it propagated from 6 $R_\odot$ to 54 $R_\odot$. The initial and final positions of the two CMEs indicates that they were deflected toward the HCS region. Both CMEs experienced deflection from high magnetic pressure regions to the low magnetic pressure regions towards the current sheet. Figure~\ref{fig:mag_env} shows the total deflection of CMEs estimated from the GCS fit to the first and last sets of images.

It has been previously reported that during solar minima, CMEs are observed to be deflected toward the equator \citep{GP_TP:2000,Cremades:2004}. Studies by \cite{gop_mac_2009} and \cite{Shen:2011} showed that the magnetic field is responsible for CME deflection. Since during the solar minimum ambient magnetic field is bipolar in nature, most of the CMEs tend to deflect latitudinally towards the equator or HCS regions, e.g., in Event 2. This deflection is from a higher magnetic pressure (far from HCS/PIL)  region to a lower magnetic pressure region (near the HCS/PIL). Moreover, the deflection in the Event 6 also agrees with this finding because it was deflected toward the PIL from the region which was far from PIL.

Event  2 had a speed of 450 km/s and showed a latitudinal shift due to magnetic field interaction.  For this CME,  magnetic field interaction is expected to dominate the drag up to 30 $R_\odot$, as reported by \cite{sach:2017} for events with similar speeds. Interestingly, this height is very close to the height up to which this CME was tracked, i.e., 34 $R_\odot$. Therefore, for Event 2, only magnetic field interaction is expected to be dominant in the domain of CME tracking, which suggested a latitudinal deflection. Therefore, the solar wind interaction is not expected to play a significant role in its propagation at these heights.

We investigated the solar wind environment of Event 6 and found that it did not influence the CME propagation. Since Event 6 is a fast CME, it is expected that drag will dominate at lower heights. Therefore, the latitudinal deflection observed after 6 $R_\odot$ (on the fully developed CME feature) can also occur due to the magnetic field interaction even below the height of the first GCS reconstruction. From GCS reconstruction in the second independent fitting at lower heights, we found the latitude was higher for  Events 2 and 6 at this time. This suggests that these CMEs had undergone a larger latitudinal deflection before the CME was well-developed in the COR2 FOV. This kind of behavior was not observed for events that showed deflection in longitude, i.e., their latitude and longitude remained unchanged compared to the original GCS fitting on well-developed CMEs  ($\approx$ 6 $R_\odot$).  Therefore, Event 6, despite being a fast event, retained its momentum for deflection in latitude from below 6 $R_\odot$. However, this was not the case with the other faster events with similar speeds, i.e., Events 1, 3, 13  and 14.

\subsection{Events showing tilt change/CME rotation}
\label{sec:Oct5}

In this section, we discuss the CME on May 23, 2017 (Event 12), which showed a change in tilt during its propagation in the heliosphere. This CME was relatively slow as compared to the other faster Events 1, 3, 13, and 14, with an average velocity of 400 km/s, which showed longitudinal deflection in the heliosphere. We reconstructed the CME from 8:24 UT on May 23 in COR2 FOV to 10:30 UT on May 24 in HI1 FOV. We did not find any significant change in the direction of the propagation of the CME as it propagated from 10 $R_\odot$ to $70 R_\odot$. However, its tilt increased from $13^\circ$ at 10 $R_\odot$ to $32^\circ$ at 70 $R_\odot$ as it propagated in the heliosphere.\\

\begin{figure}[!h]
    \begin{center}
    \includegraphics[width=17cm]{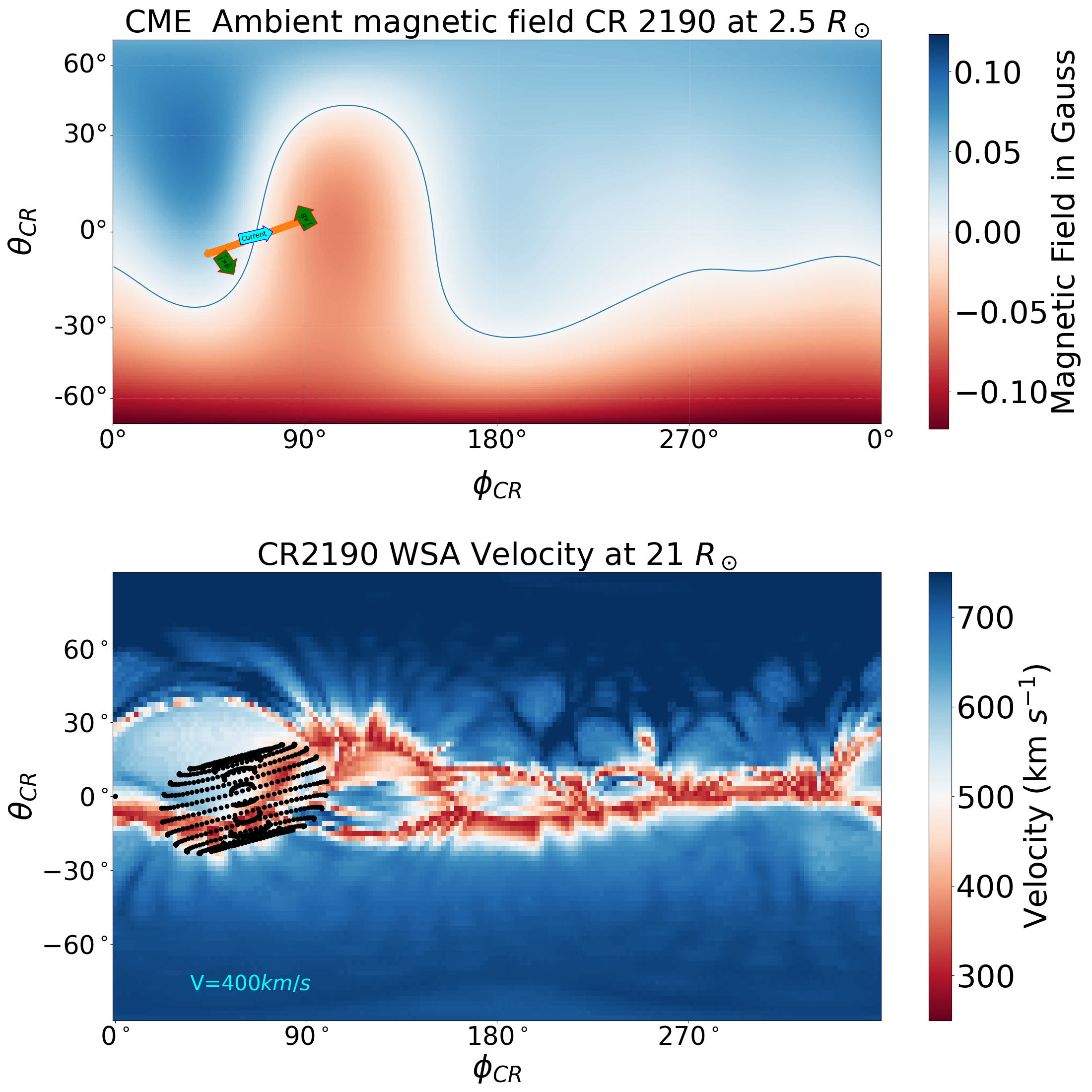}
    \caption{Radial magnetic field (top) and solar wind velocity (bottom) environment of the May 23, 2017, CME (Event 12).}
    \label{fig:Compare}
    \end{center}
\end{figure}

We investigated the ambient magnetic field environment of the CME and found that the direction of propagation is already aligned with the PIL/HCS, i.e., the center of the CME axis shown in the top panel of Figure:~\ref{fig:Compare}. Moreover, there is no noticeable difference between the velocity of the CME (400 km/s) and the ambient solar wind. Therefore, we do not expect a significant deflection assuming that deflection occurs due to magnetic pressure difference or due to drag between ambient medium and CME, i.e., no change in the $\theta$ and $\phi$ of GCS parameters \citep{Wang:2004,Shen:2011}. It was also evident from the values of latitude and longitude of CME estimated from GCS reconstruction, i.e., we did not find any significant change in the direction of the propagation of the CME.\\

\begin{figure}
    \centering
    \includegraphics[width=1.0\linewidth]{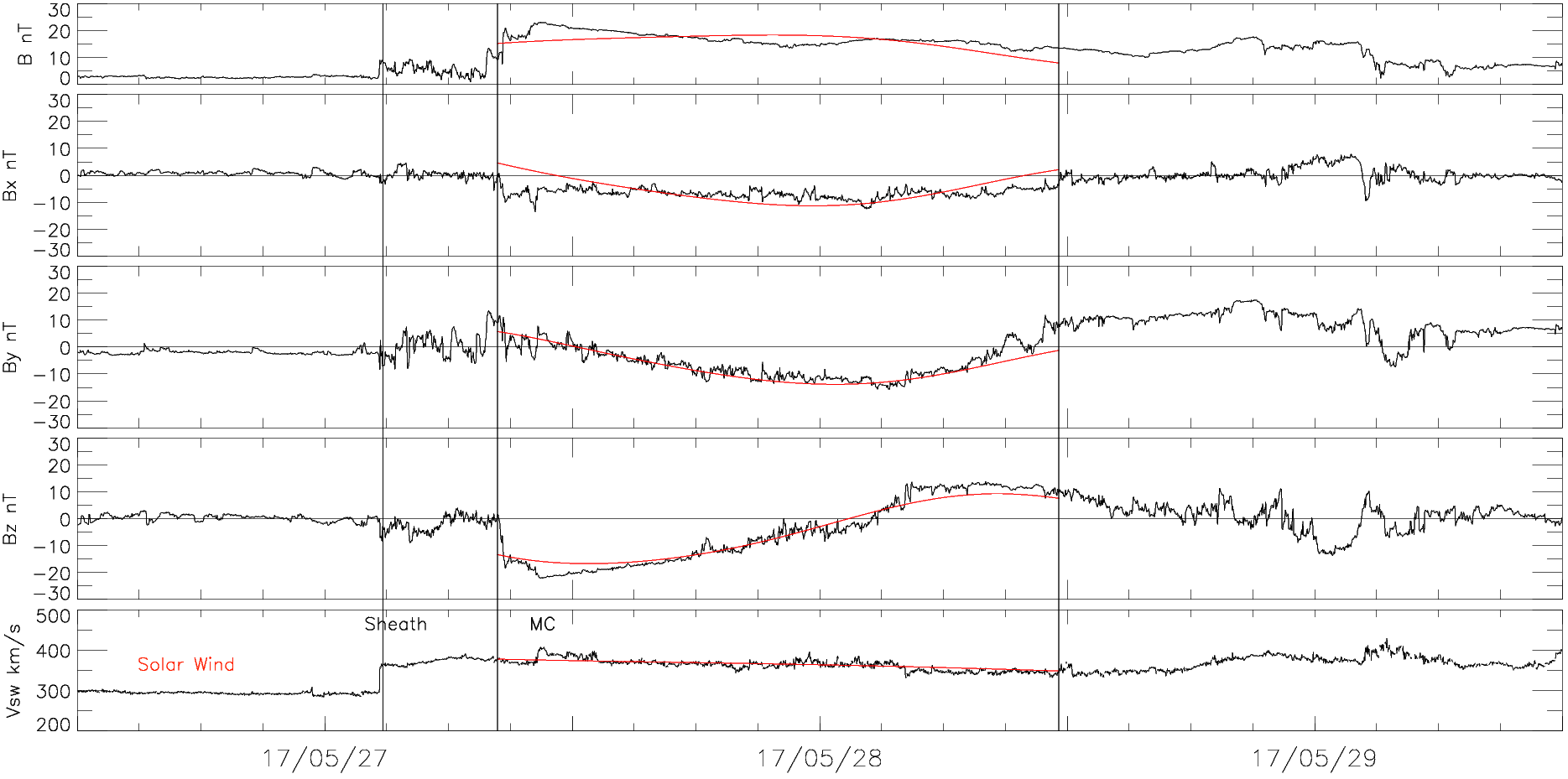}
    \caption{ Marubashi fitting of the toroidal model with solar wind parameters in the magnetic cloud (red line) and observed solar wind parameters from ACE plotted with respect to time (YY/MM/DD), i.e., B, $B_x$, $B_y$, $B_z$, of IMF and bulk solar wind velocity in km/s from top to bottom respectively (black curve).}
    \label{fig:imf_vec}
\end{figure}

Using Marubashi toroidal and cylindrical fitting models \citep{Maru:2007,Maru:2017}, we found a low inclination for the flux rope at L1 for May 23, 2017, CME. It is also evident from the rotation of the Bz in the IMF vectors with prolonged $B_y$ shown in Figure:~\ref{fig:imf_vec}, i.e., the signature of the SWN flux rope. Its tilt at the L1 was estimated $\approx 40^\circ$ which is similar to that estimated from the last GCS reconstruction in HI1 FOV, i.e., $\approx 32^\circ$. It suggests that this CME rotated by a smaller angle beyond 70 $R_\odot$, i.e., from $32^\circ$ to $40^\circ$ .\\

\begin{figure}[!h]
    \begin{center}
    \includegraphics[width=17cm]{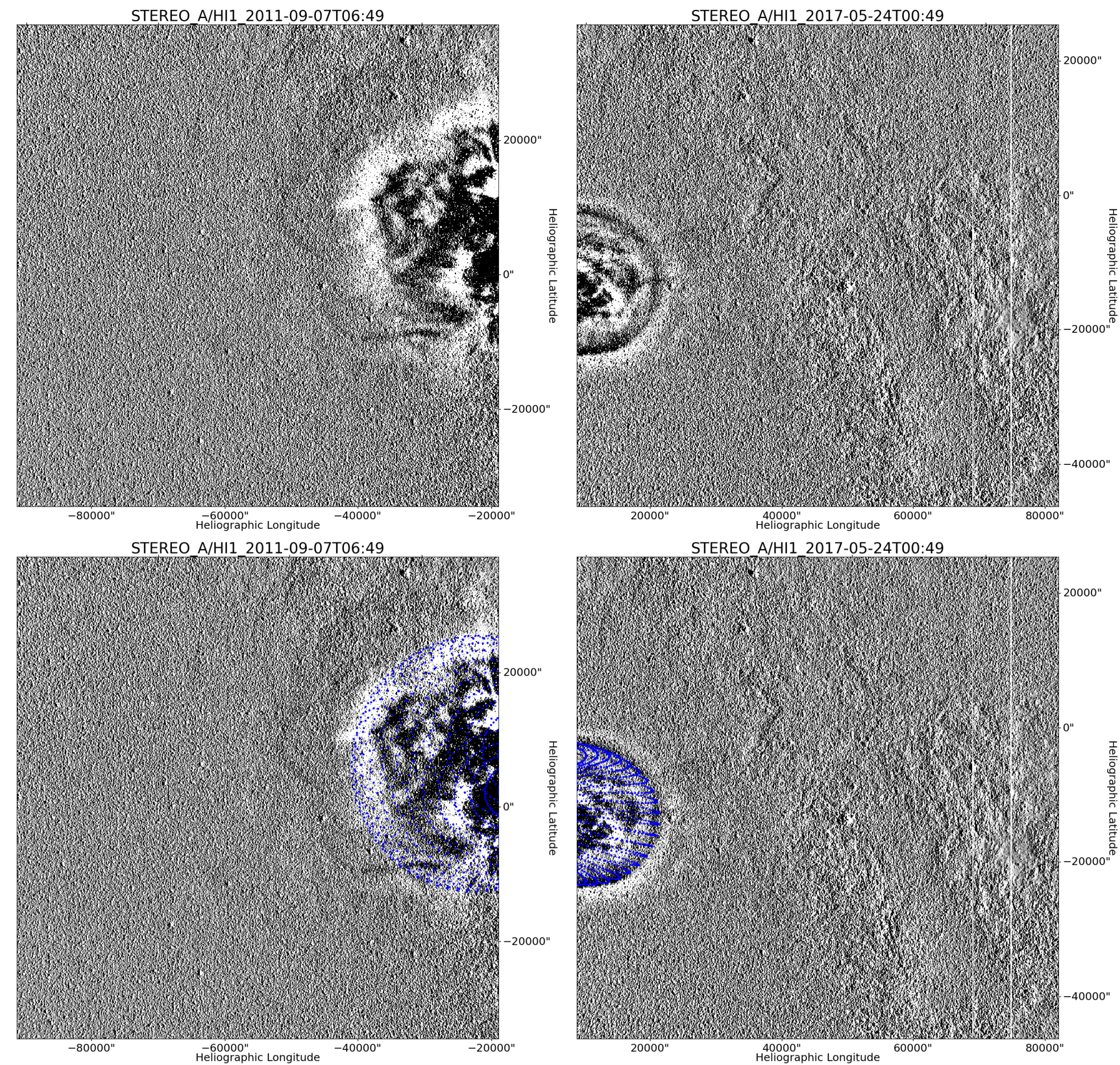} %

    \caption{ September 7, 2011 CME (Event 3) with velocity $\approx$850 km/s on the left showing a disturbed noncoherent structure at the heliocentric distance of $\approx$ 37 $R_\odot$ estimated from GCS reconstruction. May 23, 2017, CME (Event 12) on the right with velocity $\approx$ 400 km/s  showing a coherent circular structure at $\approx$ 48 $R_\odot$ in HI1 FOV. The top panel shows the CME structure without the GCS mesh, and the bottom panel shows the CME with GCS mesh over-plotted on it (one can clearly notice the differences in the quality of fit between the two cases)}.
    \label{fig:Coherent}
    \end{center}
\end{figure}

We investigated the ambient magnetic field using the PFSS model for May 23, 2017 CME to understand the role of the magnetic field in its propagation. The top panel of Figure:~\ref{fig:Compare} shows the ambient magnetic field for May 23, 2017 CME. As we can see, on the left of the PIL, the magnetic field is directed radially outward in the blue color region, and it is directed radially inward in the red region. Based on the in situ observations of IMF, we also identified the SWN type of flux rope associated with the CME. The direction of the current in this type of flux rope is from south to north, as shown by the cyan color arrow along the axis of the CME. Green arrows at the end show the direction of torque on the CME. Bottom panel of Figure:~\ref{fig:Compare}, shows the solar wind velocity background for May 23, 2017 CME. This CME is much less spread in the $\theta-\phi$ plane,  i.e., half angle $22^\circ$, and there is a smooth, slow, and fast solar wind velocity ahead of the CME front. \\
In our previous study of the October 5, 2012, CME (Event 7), we observed a continuous rotation of the CME from $44^\circ$ at around 15 $R_\odot$, approximately $65^\circ$ at 58 $R_\odot$ and $110^\circ$  at L1. This rotation was attributed to two factors: the $\vec{J} \times \vec{B}$ torque (Lorentz force) within the flux rope, particularly below 21 $R_\odot$, and interaction with the solar wind above this boundary.
The CME was an extended structure in the $\theta$-$\phi$ plane with a half angle of $52^\circ$. Figure 5 in \cite{Kumar_2023} showed distinct regions of slow and fast solar wind ahead of the CME. A non-uniform velocity profile was noted ahead of the leading edge of the fast CME, with slow solar wind preceding the upper segment and fast solar wind preceding the lower part. Interaction between the slow solar wind and the fast-moving upper segment of the CME caused an eastward deflection due to solar rotation, while the lower segment of the CME faced predominantly fast solar wind. The interaction with the ambient magnetic field and solar wind gave the CME a consistent sense of rotation throughout the heliosphere.
 \\ 

 The direction of current ($ \vec{J}$ ) in the May 23, 2017, CME  and ambient magnetic field ($ \vec{B}$) were both reversed near the Sun in comparison to the October 5, 2012, CME (Bottom panel of Figure:2, in \cite{Kumar_2023}). Therefore we found a consistent sense of $\vec{J} \times \vec{B}$ as shown in top panel of Figure:~\ref{fig:Compare} for both the events, i.e., counterclockwise.\\
 
 May 23, 2017, CME is a much less spread in the $\theta-\phi$ plane as compared to October 5, 2012, CME,   i.e., lower half angle ($22^\circ$), and there is a homogenous/mixed solar wind background ahead of the CME front. Moreover, there is no significant difference between the velocity of the CME (400 km/s) and the ambient medium solar wind velocity. Therefore, we do not expect the same kind of solar wind interaction in the May 23, 2017 CME as observed in the October 5, 2012 CME, i.e., no rotation due to solar wind interaction. Since the solar wind drag operates throughout the heliosphere, we expect the rotation of Event 7 due to drag to be long-lasting as compared to Event 12, which rotated only because of the magnetic field interaction. \\
 
In a recent study, \cite{martinic:2023}  estimated the drag parameters ($\gamma$) for the October 5, 2012, CME and May 23, 2017, CME as, 0.502 and 0.065, respectively using the reverse drag technique (DBEMv3 tool, \cite{Inv:2021}). The $\gamma$ parameter in their model is directly proportional to the cross-sectional area of the CME in the heliosphere. Also, the cross-sectional area increases with kappa and half angle in the GCS model. Therefore, our results are in agreement with \cite{martinic:2023} as we obtained a higher half angle and kappa for the October event ($52^\circ$, 0.48) than that of the May 23, 2017 event ($22^\circ$, 0.38). 
A lower value of drag parameter for May 23, 2017, CME compared to October 5, 2012, CME provides evidence of a weaker drag interaction of May 23, 2017, CME with the solar wind as compared to October 5, 2012, CME. This is confirmed by the higher change in the tilt of October 5, 2012, CME compared to May CME while propagating in the drag-dominated regime.

Therefore, for the May 23, 2017, CME, only the magnetic field environment was in favor of its rotation whereas for October 5, 2012, both the factors, i.e., magnetic field and ambient solar wind velocity, were in favor of its rotation.

It is worth mentioning that the overall shape of the two CMEs that rotated remained consistent and coherent throughout their propagation, which made it possible to implement GCS reconstruction on the CME structure and to estimate the GCS model parameters with less ambiguity. This contrasted with the other faster events analyzed in this paper, where such consistency and coherency in CME structure in HI1 images was lacking ( Figure:~\ref{fig:Coherent}).
Furthermore, we observed that the fast CMEs in our dataset were fitted with larger values of kappa, this agrees with previous studies by \cite{pluta:2019}. As kappa increased, we encountered difficulty in distinguishing the quality of GCS fits with different tilts on the images. It became challenging to differentiate the projection of the GCS mesh with different tilts for those CMEs that were fitted with higher kappa values.\\
\section{Conclusions}
An analysis of 15 geo-effective CMEs observed by SOHO/LASCO/C2 \& C3, STEREO/SECCHI COR2, and HI1, reveals that CME propagation can be influenced by ambient magnetic field and solar wind velocity, causing deviations from self-similar expansion. \\

Our analysis shows the latitudinal deflection predominantly occurred in the equatorward direction toward the HCS in two events (Events 2 and 6). We found the net latitudinal deflection observed (from the first COR2 frame to the last frame in HI1), for these events, is toward the PIL and is consistent with the results of \cite{Shen:2011}. It is important to mention here that their results were based on tracking of the CMEs up to LASCO/C3 FOV only. We believe that the deflection observed towards the HCS beyond 21 $R_\odot$ might be either due to the earlier momentum gained by the deflection in the magnetically-dominated regime (below 21 $R_\odot$) or due to interaction with the solar wind or both.\\

Interestingly, faster CMEs (Event 1, 3, 13 and 14) in our study, when propagating away from the Sun-Earth line in a westward direction, were deflected towards the east, i.e., toward the Sun-Earth line by the slower solar wind ahead of them. This is also in agreement with results reported in earlier studies \citep{Wang:2004,Wang:2014}. Moreover, we observed a dependence of the magnitude of longitudinal deflection on the velocity of the CMEs.\\    
  Our analysis revealed a change of $19^\circ$  in the tilt of May 23, 2017, CME in the heliosphere. Notably, this rotation is similar to that we reported recently in \cite{Kumar_2023}, where a $21^\circ$ rotation in the heliosphere from 14 $R_\odot$  to 58 $R_\odot$ was reported. It is important to note that although the magnetic field configuration for May 23, 2017, CME favored an anti-clockwise rotation in the CME flux rope, however the increase in tilt/CME rotation persisted even beyond the magnetically dominated regime, i.e., from 21 $R_\odot$ to 70 $R_\odot$. We attribute this change to potentially gained angular momentum below 21 $R_\odot$. It is worth pointing out that the height ($\tilde{h}_0$) at which the drag starts to dominate the magnetic interaction (Lorentz force) varies for different CMEs \citep{sach:2015,sach:2017,martinic:2023}. \cite{martinic:2023} concluded that the range of $\tilde{h}_0$ can be  from 3.5 $R_\odot$  to 70 $R_\odot$. This range also defines the upper bound of the height up to which magnetic field interactions can influence the CME trajectory. For slower CMEs (speed$\leq$ 900 km/s), this height tends to be higher compared to faster CMEs. In particular, for October 5, 2012, CME, the height $\tilde{h}_0$ was estimated as 31  $R_\odot$ by \cite{sach:2017}. Although their dataset did not include the May 23, 2017 event, however, they found $\tilde{h}_0$ is up to $\approx$ 47 $R_\odot$ for CMEs with comparable speeds to the May 23, 2017 CME (Refer to Tables 1 and 2 in \cite{sach:2017}). The above observations strongly support the idea that CME rotation due to Lorentz force is possible even beyond 21 $R_\odot$  for these two CMEs. 
 
 However, the analysis showed that for May 23, 2017, CME rotation did not persist in the drag-dominated regime for a longer time, i.e., beyond 70 $R_\odot$. This is evident from the estimated value of tilt at 70 $R_\odot$ ($\approx32^\circ$)  from GCS reconstruction and at L1 ($\approx 40 ^\circ$) from Marubashi fitting. This contrasts the behavior observed for October 5, 2012, CME, where a continuous rotation from 14 $R_\odot$ to L1 was identified.\\

 A recent simulation study by \cite{Koehn:2022} showed that the orientation of the CME flux rope affects its geoeffectiveness. Keeping all the parameters constant in a spheromak model of a CME, they found that a tilt of $180^\circ$ of the spheromak axis (equivalent to $90^\circ$ tilt of the CME flux rope as defined in our study) results in a prolonged southward $B_z$, leading to the most geoeffective case in their study. In our study, the high inclination ( October 5, 2012 event) was found to be less geoeffective than the low inclination CME (May 23, 2017) because they had different values of $|B|$, which manifested in different values of $B_z$.

 We want to point out that in our current work, we only focused on geoeffective CMEs. There is a need to analyze more events to further study phenomenon of rotation in CMEs in details and their impact.\\
 Our study, based on 15 events, also shows that CME rotation is a rare phenomenon observable in HI1 images. This rarity can be partially attributed to the special conditions required for CME rotation. Although CME rotation appears to be more commonly observed in the lower corona where the ambient magnetic field dominates, it requires conducive conditions of both the magnetic field and the solar wind to favor a persistent rotation of the CME throughout the heliosphere as recently reported in \cite{Kumar_2023}. This, combined with the fact that CME becomes fainter and incoherent as it reaches the HI1 FOV, makes it difficult to implement GCS fitting. The above-mentioned factors make CME rotation a rare phenomenon to observe in the heliosphere.\\

Furthermore, we believe that we were able to capture the total change in the orientation of the CME (i.e., tilt) for two CMEs in our data set of 15 CMEs because we could track these CMEs in the heliosphere over a larger distance, compared to previous studies that focused on stereoscopic reconstruction up to LASCO/C3 FOV only. This further suggests that HI1 observations help us to connect observations near the Sun and near Earth, improving our understanding of how CMEs move through the heliosphere.\\
\begin{acknowledgments}
We thank the reviewer for the constructive comments to improve the manuscript. We thank and acknowledge K. Marubashi for providing us with the flux rope fitting code in IDL. We also acknowledge the use of ACE magnetic field and velocity data (\url{https://izw1.caltech.edu/ACE/ASC/level2/lvl2DATA_MAG.html}) and GONG program for the synoptic map data (\url{https://gong.nso.edu/data/magmap/crmap.html}.). We acknowledge the use of the GCS and $pfsspy$ code. We acknowledge the Sunpy community\citep{Sunpy_community2020}. We acknowledge the NASA/CCMC program for providing us the WSA velocity map at 21 $R_\odot$. Coronagraphic images are used from Helioviewer and HI1 images from the site (\url{https://stereo-ssc.nascom.nasa.gov/data/ins_data/secchi/secchi_hi/L2_11_25/}). This work was carried out under the Indo-U.S. Science and Technology Forum (IUSSTF) Virtual Network Center project (Ref. no. IUSSTF/JC-113/2019). NG is supported by NASA's STEREO project and Living With a Star program. Ashutosh Dash acknowledges the opportunity provided by PRL to carry out his final year master's project at USO/PRL. 
\end{acknowledgments}

\bibliography{manuscript}{}
\bibliographystyle{aasjournal}

\end{document}